\documentstyle[fleqn]{article}
\newcommand{\bg}{\begin{equation}}
\newcommand{\eg}{\end{equation}}
\def\diag{\mbox{diag }}
\def\ZZ{\mbox{\rm Z\hskip-.35em Z}}

\def\NN{\mbox{\rm I\hskip-.15em N}}

\def\CC{\mbox{\rm C}\hskip-.5em\mbox{l} \;}
\def\id{\mbox{\rm 1\hskip-.25em l}}
\def\HH{\mbox{\rm I\hskip-.15em H}}

\def\dd{\mbox{d}}
\def\tr{\mbox{tr }}

\def\M{^{\mu}}

\def\LD{{\cal L}}

\def\AL{{\cal A}}

\def\lplus{\supset\mkern -19mu\hbox{\small +}\mkern 
10mu}

\newcommand{\mb}[1]{\mbox{\boldmath $#1$}}
\newcommand{\mr}[1]{\mbox{\bf #1}}
\begin{document}
\begin{center}
{\Large THE STANDARD MODEL WITHIN\\NONCOMMUTATIVE 
GEOMETRY:\\[3pt]A COMPARISON OF MODELS}
\end{center}
\vspace*{2cm}
\begin{center}
 Florian Scheck\\
Institut f\H ur Physik - Theoretische 
Elementarteilchenphysik,\\
Johannes Gutenberg-Universit\H at\\ D-55099 Mainz
\end{center}
\vspace*{3cm}
\begin{abstract}
Algebraic Yang-Mills-Higgs theories based on 
noncommutative geometry have
brought forth novel extensions of gauge theories with 
interesting
applications to phenomenology. We sketch the model of 
Connes and Lott,
as well as variants of it, and the model developed by a 
Mainz-Marseille
group, by comparing them in a schematic way. The role of 
fermion masses
and mixings is discussed, and the question of possible 
parameter relations
is briefly touched.
\end{abstract}
\vfill
\noindent
\rule[.1in]{14cm}{.003in}
Contribution to IX-th Max Born Symposium, Karpacz, 25 - 
28 September 1996
\newpage
\noindent
{\bf 1. Introduction}\\

Application of ideas of noncommutative geometry 
\cite{Con} to particle physics
has brought forth, over the last couple of years, novel 
and phenomenologically
interesting extensions of Yang-Mills theories 
\cite{CoL,CEV,CES,HPS,CHPS}, 
most of which
were developed simultaneously. In fact, there is by
now a problem of nomenclature because people have become 
used to associate
noncommutative versions of the standard model with the 
specific framework proposed
by Connes and Lott \cite{CoL} although there are 
alternatives whose construction
principles and whose results, with regard to physics, 
are  different. 
One of the purposes of 
this lecture is to make a comparison between two of 
these approaches
both of which have the virtue of having been elaborated 
in some detail. 
Therefore, in order to avoid that confusion I propose to 
call the class of extended 
Yang Mills theories using ideas
of noncommutative geometry generically {\em algebraic 
Yang-Mills-Higgs theories\/},
independently of whether or not they follow Connes' 
specific constructive
framework.

In sec.~2 I shortly describe the Connes-Lott 
construction, including a 
variant of it \cite{Wul}, 
and the model proposed by a Mainz-Marseille group, 
sketch the comparison
between them, work out similarities and point out 
salient differences. 
Sec.~3  addresses the role of fermion masses and fermion 
state mixing in these 
models. In sec. 4 I comment on possible relations 
between parameters of the fermion 
sector that are claimed to follow from the Connes-Lott 
model.\\[4pt]

\noindent
{\bf 2. Algebraic Yang-Mills-Higgs theories}\\

I shall describe primarily the model proposed by Connes 
and Lott \cite{CoL} including 
a variant of it that was proposed by the Leipzig group 
\cite{Wul}, and the 
Mainz-Marseille model but much of what will be said also 
applies to other
constructions. Let me begin by listing the {\em common 
features\/}:

(i) What the models have in common, at the algebraic 
level, is a
differential algebra,
$
\left(\Omega^\ast ,\dd\right)\, ,
$
which may be $\NN$-graded or $\ZZ_2$-graded. The algebra 
is
equiped with a product, in most cases associative, which 
is not
the same in different 
models. In an appropriate limit applied to the algebra, 
one recovers
the de~Rham algebra
$\left( \Lambda^{\ast} ,\dd_C \right)_{\NN}$, with 
$\dd_C$ the
ordinary Cartan exterior
derivative, the product being the familiar wedge 
product. 
That is to say, one starts from an algebra $\AL$, 
commutative or
noncommutative, replacing $C^\infty (M)$, the algebra of 
smooth functions over
Minkowski space $M$, and constructs a differential 
algebra $\Omega^\ast(\AL )$ 
from it, with, of course, $\Omega^0=\AL$. 

(ii) At grade one the space $\Omega^1$ is
often the same so that the generalized connection 
$\mr{A}$ is also the same.
However, the differentials $\dd$ and, hence, 
the generalized curvature $\mr{F}$ differ.

(iii) The generalized connection $\mr{A}$ incorporates 
the ordinary, spin-1,
gauge boson fields, say $V$, $W$, etc, and the Higgs 
multiplet(s), say $\Phi$.
 
(iv) The action (usually)
exhibits spontaneous symmetry breaking, the Higgs coming 
about rather naturally
as a geometric phenomenon. For example, in the 
Mainz-Marseille model there is a
direct link between spontaneous symmetry breaking in the 
electroweak sector
and maximal parity violation of charged weak 
interactions - a most remarkable
feature.

The {\em differences\/} between different constructions, 
in short, 
may be grouped as follows.

(a) The role of the Dirac operator is rather different 
in different constructions.
In the Connes-Lott model, following the very spirit of 
Connes' noncommutative 
geometry, the Dirac operator is a fundamental ingredient 
that determines the
metric properties. In the Mainz-Marseille model the 
Dirac operator is a derived
quantity. One first constructs the bosonic sector of the 
standard model by
itself, then incorporates the fermions in chiral 
representations.

(b) The explicit form of the Higgs potential comes out 
different, at the
classical level, and I will comment on this difference 
below.

(c) The assignment of quantum numbers of quarks and 
leptons needs some fixing
in the Connes-Lott approach while it comes out right, 
without further input,
in Mainz-Marseille. Obviously, this is important in the 
context of cancellation
of anomalies.

(d) The limit of ''vanishing noncommutative structure'', 
in the two classes of
models, is technically different. While it is obvious in 
the Mainz-Marseille
model it is less transparent in Connes-Lott.

(e) The interpretation in terms of the underlying 
geometry, although it is not
fully understood as yet, may be quite different. Connes' 
approach is
the deeper and more ambitious one because, qualitatively 
speaking, it aims at
replacing the geometry of spaces by the study of 
algebras. For this reason it
lends itself rather naturally to inclusion of gravity in 
constructing actions
-- a feature foreign to the more pragmatic 
Mainz-Marseille model. The latter,
on the other hand, has the virtue of assigning a genuine 
dynamic
role to fermion fields of definite chirality, in accord 
with what the
phenomenology of electroweak interactions keeps telling 
us.\\[2pt]

\noindent
2.1 The Connes-Lott approach

This approach is based on a spectral triple $({\cal 
A},{\cal H},D)$, with $\cal A$
a unital star algebra, $\cal H$ a Hilbert space (of 
fermionic states), and $D$ a
generalized Dirac operator. From this triple one 
constructs first the universal
differential envelope $\left(\widetilde{\Omega}^\ast 
({\cal A}),\tilde{\dd}\right)$, 
and, given a representation of the algebra on $\cal H$, 
a representation of that
universal object on the Hilbert space by means of 
bounded linear operators, viz.
\begin{eqnarray}
\pi & : & \widetilde{\Omega}^\ast ({\cal 
A})\longrightarrow L({\cal H})\\\nonumber
 & & a_0\tilde{\dd}a_1\ldots\tilde{\dd}a_n\longmapsto 
\underline{a}_0[D,\underline{a}_1]\ldots 
[D,\underline{a}_n]\; .
\end{eqnarray}
(The representation of the element $a_i$ of the algebra 
on $\cal H$ is denoted 
$\underline{a}_i$.) For the construction of the standard 
model 
the Dirac operator, very schematically, will have the 
form 
$D=i\gamma^\mu\partial_\mu +D_M$, the discrete piece 
being
\bg
D_M=\left(\begin{array}{cc|cc}
\mb{0} & M & \mb{0} & \mb{0} \\M^\dagger & \mb{0} & 
\mb{0} & \mb{0} \\\hline
\mb{0} & \mb{0} & \mb{0} & M^\ast \\ \mb{0} & \mb{0} & 
M^T & \mb{0}
\end{array}\right)\, ,
\eg
where $M$ is the fermionic mass matrix, i.e.
\bg
M=\left(\begin{array}{cccc}
M_u\otimes\id_3 & & & \\ & M_d\otimes\id_3 & & \\ & & 0 
& M_\ell
\end{array}\right)\, ,
\eg
where $M_u$, $M_d$, and $M_\ell$ denote the mass 
matrices of {\em up\/}-type quarks,
of {\em down\/}-type quarks, and of charged leptons, 
respectively, $M_d$ containing
the empirical Cabibbo-Kobayashi-Maskawa (CKM) mixing.

There is a technical difficulty at this point whose 
resolution, at the classical level,
has profound physical consequences: the projection 
$\pi$, eq. (1), fails to respect
the differential structure of $\widetilde{\Omega}^\ast$. 
(There exist elements 
$b\in\widetilde{\Omega}^\ast$ for which $\pi (b)$ 
vanishes while 
$\pi (\tilde{\dd} b)$ does not.)
Therefore, one has to divide out the ideals
\[
{\cal J}^k({\cal A})={\cal K}^k+\tilde{\dd}{\cal 
K}^{k-1}\, ,\qquad
\mbox{where }{\cal K}^k=\mbox{ker 
}\pi\cap\widetilde{\Omega}^k\, ,
\]
sometimes called the ''junk'', thus obtaining 
\bg
\Omega_{DY}^k=\widetilde{\Omega}^k({\cal A})/{\cal 
J}^k({\cal A})=
\pi\left(\widetilde{\Omega}^k({\cal A})\right) 
/\pi\left({\cal J}^k({\cal A})\right)\, .
\eg
I have given the index DY to this object, because the 
operator $D$, when 
supplemented by the gauge and Higgs fields, yields both 
the gauge and the
Yukawa interactions and, thus, should be termed 
Dirac-Yukawa operator. 

The exterior derivative, that belongs to this 
differential algebra, is defined by
eq. (1), while the product, denoted here by $\odot$, 
reflects the division
by the ''junk''.  

The choice of the algebra is suggested by the gauge 
group(s) one wishes to obtain 
and, for the purposes of the standard model, is taken to 
have the product form
\bg
{\cal A}={\cal A}_M\otimes C^\infty(M)\, ,
\eg
with ${\cal A}_M$ a block-diagonal matrix algebra which 
contains the gauge groups
as the subsets of its unitaries. In the case of the 
standard model one usually
takes
\[ {\cal A}_M=\CC\oplus\HH\oplus M_3(\CC )\, ,\]
times an appropriate number of copies of $\CC$ in order 
to keep track of the number
of generations. The first two terms contain the gauge 
group of electroweak interactions,
the last term contains the gauge group of QCD.

Thus, at the algebraic level, the resulting differential 
algebra is
\bg
\left(\Omega_{DY}^\ast({\cal 
A}),\odot,D_{DY}\right)_{\NN} \, ,
\eg
with the product $\odot$ and the differential as defined 
above.

Before I turn to the Mainz-Marseille approach I wish to 
add a further comment.
The Connes-Lott construction, as sketched here, is also 
the one followed by
\cite{Kas} and by \cite{GV} as well as by others. The 
Leipzig group has
varied the theme by using projective modules (whereby 
the algebras may be
chosen smaller than above)\cite{Leip} and/or by 
replacing the algebra by a 
Lie algebra \cite{Wul}. The first choice brings the 
Connes-Lott construction 
closer to the Mainz-Marseille one, to be described next, 
although there is 
no complete equivalence. The second choice has somewhat 
different 
phenomenological consequences on which we comment in 
sect. 4 below.
\\[2pt]

\noindent
2.2 The Mainz-Marseille model

The algebraic structure of the Mainz-Marseille 
construction can be summarized
as follows, again very schematically: Let $M_3(\CC )$ be 
the set of 
$3\times 3$-matrices, $\ZZ_2$-graded by means of the 
grading automorphism
$\Gamma =\diag (1,1,-1)$. This means that with $M\in 
M_3(\CC )$, the projections
\[ M+\Gamma M\Gamma\qquad\mbox{and}\qquad M-\Gamma 
M\Gamma \]
are the even and odd parts of $M$, respectively. The 
differential algebra
is taken to be the skew-tensor product
\bg
\Omega^\ast_{MM}(X)=M_3(\CC 
)\widehat{\otimes}\Lambda^\ast (X)\, ,
\eg
where $\Lambda^\ast (X)$ are the exterior forms on 
(flat) space-time $X$. 
The object (7) contains two gradings, the $\ZZ_2$ of the 
matrix factor and the
$\NN$-grading of exterior forms. When the latter is 
turned into a $\ZZ_2$-grading
for odd and even forms (i.e. by taking the exterior form 
grade modulo 2),
$\Omega^\ast_{MM}(X)$ inherits a $\ZZ_2$ bi-graded
structure. Its elements are form-valued matrices whose 
total grade is the sum,
modulo 2, of the form grade (even or odd) and the matrix 
grade. In other terms,
the product that goes with the differential algebra of 
this model and that we
denote by $\bullet$ here, must be
consistent with that graded structure but, obviously, 
there is nothing 
analogous to the division by the ideals of eq. (4). The 
differential $\dd_{MM}$,
finally, is chosen to be
\bg
\dd_{MM}=\dd_C + [\eta ,\,\cdot\;  ]_g\, ,
\eg
where $\eta$ is a fixed, odd element of $M_3(\CC )$, and 
where $[\,\cdot\, ,\,\cdot\, ]_g$
denotes the graded commutator. 

Thus, algebraically, the Mainz-Marseille model has the 
structure
\bg
\left(\Omega_{MM}^\ast (X),\bullet 
,\dd_{MM}\right)_{\ZZ_2}\, .
\eg
Here again I wish to add a few remarks. The action of 
$\dd_{MM}$ on an element
$M$ of $M_3(\CC )$ is defined in an essentially unique 
way \cite{CHPS}. 
Indeed, one shows that
the apparent freedom in choosing the odd element $\eta$ 
is a freedom of choosing
a phase and a ''strength''. The former corresponds to 
the freedom
of choosing the vacuum point in the degenerate Higgs 
potential, whereby all
choices are physically equivalent, the latter can always 
be absorbed in the
(only) mass scale that appears in the connection form. 
Furthermore, one sees
that the bi-graded structure described above effectively 
means embedding the 
Lie algebra of $SU(2)_L\times U(1)$ of electroweak 
interactions in the minimal 
graded Lie algebra $su(2\vert 1)$ \cite{CES,HPS}. Thus, 
in the actual calculations 
one may either use the explicit rules defined above, or 
the algebra structure
encoded in the graded commutator of $su(2\vert 1)$. The 
two methods are 
completely equivalent. Note, however, that this graded 
Lie algebra neither
represents a new symmetry of the theory nor is any 
attempt made to gauge it. 
I mention this because there was some confusion with the 
much earlier 
work of Ne'eman, Thiery-Mieg, Fairlie and others 
(see, e.g. \cite{NTM} and further references in 
\cite{CES})
who also used $su(2\vert 1)$ in the context of 
electroweak interactions. 
A closer examination of their pioneering work shows that 
the role of this algebra 
is very different here.\\[2pt]

\noindent
2.3 A toy model for comparison

It is instructive to illustrate and to compare the two 
approaches by a simple
example \cite{PPS} which does show all relevant features 
but avoids the complexity
of the full standard model. For that purpose choose the 
matrix algebra in the
Connes-Lott approach to be
$ {\cal A}_M=\CC \oplus\CC \, , $
and, in the Mainz-Marseille case, to be
$M_2(\CC )$, equipped with the grading automorphism 
$\Gamma =\diag (1,-1)$. 
The odd element $\eta$ in eq. (8) here is chosen to be
\[
\eta =i\left(\begin{array}{cc}
0 & c \\ \bar{c} & 0
\end{array}\right)\, ,
\]
with $c$ a dimensionless complex number, while the
Dirac-Yukawa operator in the
corresponding Connes-Lott construction is $D=\mu\eta$, 
with $\mu$ a
constant with dimension mass. In constructing the 
differential algebra
$\Omega_{DY}^*({\cal A})$ it is not difficult to see 
that the ''junk'',
for the first three grades, is given by
\bg
\pi ({\cal J}^0)=\{ 0\} =\pi ({\cal J}^1)\, ,\quad
\pi ({\cal J}^2)=M_0\otimes\Lambda^0(X)\, ,
\eg
with $M_0$ an even matrix, $\Lambda^0(X)$ functions on 
$X$. Furthermore,
the differential $\dd_{DY}$ can be written in a form 
analogous to (8).
In essence, this means that
\[
\Omega_{DY}^2\left( {\cal A}_M\otimes C^\infty 
(X)\right)\cong
\left(\begin{array}{cc}
\Lambda^2 & 0 \\ 0 & \Lambda^2
\end{array}\right)+
\left(\begin{array}{cc}
0 & \Lambda^1 \\ \Lambda^1 & 0
\end{array}\right)\, ,
\]
and that in calculating products $\odot$ and 
differentials $\dd_{DY}$,
all terms of the form
\[
\left(\begin{array}{cc}
\Lambda^0 & 0 \\ 0 & \Lambda^0
\end{array}\right)
\]
are dropped. Note that this truncation does not happen 
in the
Mainz-Marseille case.

In either model the generalized connection is the same 
(at grades 0
and 1 the spaces $\Omega^{k}$ coincide). It reads, in 
appropriate units,
\bg
\mb{A}=i\left(\begin{array}{cc}
V & \Phi \\ \Phi^* & W \end{array}\right)\, .
\eg
At grade 2, however, the models differ, for the reasons 
explained above.
In the Connes-Lott case the generalized curvature of the 
toy model is
\begin{eqnarray}
\mb{F}^{(CL)} & = & \dd_{DY}\mb{A}+\mb{A}\odot\mb{A}\\ 
\nonumber
 & = & i\left(\begin{array}{cc}
\dd_CV & -\dd_C\Phi-i(V-W)(\Phi+1)\\
-\dd_C\Phi^*+i(V-W)(\Phi^*+1) & \dd_CW
\end{array}\right)\, .
\end{eqnarray}
Obviously, in calculating the Lagrangian from (12) one 
will find
\bg
\LD^{(CL)} 
=-\frac{1}{4}F^{(V)\;2}-\frac{1}{4}F^{(W)\;2}+
2\left( \mr{D}\Phi\right)^*\left( \mr{D}\Phi\right)\, ,
\eg
i.e. an expression that does contain the correct 
covariant derivative
of the Higgs field
\[
\mr{D}\Phi =\dd_C\Phi +i(V\Phi -\Phi W)+i(V-W)\, ,
\]
but no Higgs potential.

The same calculation for the Mainz-Marseille case gives 
a different
result for the curvature, viz.
\begin{eqnarray}
\mb{F}^{(MM)} & = & \dd_{MM}\mb{A}+\mb{A}\bullet\mb{A}\\ 
\nonumber
 & = & i\left(\begin{array}{cc}
\dd_CV -(\Phi +\Phi^*+\Phi\Phi^*) & 
-\dd_C\Phi-i(V-W)(\Phi+1)\\
-\dd_C\Phi^*+i(V-W)(\Phi^*+1) & \dd_CW-(\Phi 
+\Phi^*+\Phi\Phi^*)
\end{array}\right)\, .
\end{eqnarray}
The Lagrangian obtained from this expression reads
\bg
\LD^{(MM)}=\LD^{(CL)}+U(\Phi )\, ,\mbox{ with}
\eg
\bg
U(\Phi )=2\left(\Phi +\Phi^* +\Phi\Phi^*\right)^2\, .
\eg
Note that it exhibits spontaneous
symmetry breaking, the potential $U(\Phi )$
stemming from the new, derivative-free terms in the 
diagonal of $\mb{F}$.
(The Higgs field appears here in a dimensionless form. 
The physical field
is related to it by a dimensionfull scale factor.) While 
the gauge field
'$\gamma$'$=V+W$ remains massless, the field '$Z$'$=V-W$ 
becomes
massive, the mass term stemming from the off-diagonal 
terms in $\mb{F}$.
The original gauge symmetry $U(1)\times U(1)$ is broken 
to the residual
symmetry $U(1)_{\gamma}$.

The occurence of the Higgs potential (16) is very 
natural in the context
of these models. Indeed, in the case of the
Mainz-Marseille model spontaneous
symmetry breaking (SSB) is a direct consequence of the 
noncommutative
differential structure encoded in (8): If one replaces 
$\eta$
by $\rho\cdot\eta$ where $\rho\in [0,1]$ is a parameter 
that
is introduced temporarily to control the ''amount of 
noncommutativity''
in the model, one sees that for $\rho =0$ there is no 
SSB, both gauge
fields remain massless, and the model keeps its full 
gauge symmetry.
The reason for the absence of SSB in the Connes-Lott 
construction of
the toy model is to be found in the subtraction of the 
''junk''. In
this example taking out the ideal $\pi ({\cal J}^2)$, 
eq. (10), is
equivalent to dropping the term $(\Phi 
+\Phi^*+\Phi\Phi^*)$ in the
diagonal of (12).\\[2pt]

\noindent
2.4 The standard model case

The Mainz-Marseille construction of the electroweak 
sector of the
standard model is fairly straightforward. As is was 
described elsewhere
in quite some detail \cite{CEV,CES,HPS,CHPS} we just 
summarize the
results here. The generalized connection has the form
\bg
\mb{A}=i\left(\begin{array}{c|c}
(a/2)\mb{\tau}\cdot\mb{W}+(b/2)\id_2W^{(8)} & (c/\mu 
)\overline{\mb{\Phi}} \\
\hline (c/\mu )\mb{\Phi} & bW^{(8)}
\end{array}\right)\, ,
\eg
where $\mb{\Phi}=\left(\Phi^{(0)},\Phi^{(+)}\right)^T$ 
now is the Higgs
doublet, while the assignment of the gauge fields is 
obvious. The
resulting Lagrangian as obtained by means of eqs. (7) - 
(9) is precisely
the one of the standard model with SSB, the neutral 
Higgs field appearing
in the correct, shifted, phase \cite{CHPS}. 

The Higgs phenomenon obtains
a simple and transparent interpretation: A closer 
examination of the gauge
transformations acting on $\mb{A}$ shows that there is 
one constant
connection which is invariant under all global gauge 
transformations.
The corresponding generalized curvature is found to be
\bg
\mb{F}_{bg}^{(MM)}=i\left( I_3+Y/2\right)\equiv 
Q_{e.m.}\, .
\eg
Thus, the model is characterized by a constant {\em 
background field,\/} in the
internal symmetry space, which is nothing but the charge 
operator. This
explains at once why the residual symmetry is the $U(1)$ 
of
electromagnetism.

The analogous Connes-Lott construction also leads to the 
standard model
Lagrangian (including QCD) with the electroweak sector 
in the spontaneously
broken phase. There are important differences, however. 
The most
pronounced difference occurs in the Higgs potential 
$U(\mb{\Phi})$
because the two parameters that determine $U(\mb{\Phi})$ 
are functions
of the empirical fermionic mass and mixing matrix (2). 
In particular,
as a direct consequence of subtracting the ''junk'', cf. 
eq. (4),
the potential is proportional to the square of the 
perpendicular part
of $MM^\dagger$,
\bg
\tr \left(MM^\dagger \right)_{\perp}^2 \, ,
\eg
where the perpendicular part of $MM^\dagger$ is defined 
as follows
\[
\left(MM^\dagger \right)_{\perp}:=MM^\dagger -
\id_n\frac{1}{n}\tr \left(MM^\dagger \right)\, ,
\]
with $n=\mbox{dim }M$. Obviously, the trace (19) 
vanishes if and only if
$\left(MM^\dagger \right)$ is proportional to the unit 
matrix.
In the case of the standard model, this happens either 
if there is only
one generation, or if, in each charge sector, the 
generations are
degenerate in mass. Thus, at the classical level, there 
is SSB only
if there is more than one generation. The parameters of 
the
electroweak sector are functions of the empirical 
fermion masses -
in marked contrast to the Mainz-Marseille model whose 
bosonic sector is
independent of the fermions in the theory.\\[4pt]

\noindent
{\bf 3. Role of fermion masses and mixing}\\

In the Connes-Lott approach the fermionic mass and 
mixing matrices (3)
are taken as the essential {\em input\/} of
the Dirac-Yukawa operator. This is both the strength and 
the weakness
of the model: The empirical values of the fermion masses 
and of
their mixing matrix elements, as determined from 
experiment, fix to
a large extent the bosonic sector of the standard model. 
Of course,
with these parameters interpreted as being fundamental 
there will be
no possibility for ever calculating them or relating 
them to each other
or to other parameters of the theory.

This is not so in the Mainz-Marseille construction. In 
this model the
bosonic sector is obtained from the differential algebra 
(9) and, hence,
lives on its own, independently of whether or not there
are fermions in the theory. Furthermore, the model to 
some extent
allows to relate masses and mixings in a physically 
plausible way
\cite{CES,HS}. This is due to the specific algebraic 
sructure of the
model. Indeed, and as we mentioned above, the 
differential algebra~(9)
carries the bi-graded structure of $su(2\vert 1)$ which, 
therefore,
pops up as a kind of classifying algebra for the bosonic 
sector.
It then seems natural to classify also the fermions by 
means of
representations of $su(2\vert 1)$. The consequences of 
this hypothesis
are interesting and I briefly sketch them here.

It was well known from the work of Ne'eman and 
Thiery-Mieg \cite{NTM}
that quarks and leptons fit perfectly into the 
fundamental representations
of $su(2\vert 1)$. For example, for one generation of 
leptons and of
quarks, respectively, we have
\bg
\rho^{(\ell)}=[I_0=\frac{1}{2},Y_0=-1]  \longrightarrow 
(\frac{1}{2},-1)\oplus (0,-2)\, , 
\eg
\bg
\rho^{(q)}=[I_0=\frac{1}{2},Y_0=\frac{1}{3}]  
\longrightarrow 
(\frac{1}{2},\frac{1}{3})\oplus (0,\frac{4}{3})\oplus
(0,-\frac{2}{3})\, ,
\eg
where the right-hand side gives the decomposition in 
terms of multiplets
$(I,Y)$ of $SU(2)\times U(1)$. Also, a singlet, 
right-handed neutrino
would be described by the trivial representation
\bg
\rho_0^{(\ell)}=[I_0=0,Y_0=0]\longrightarrow (I=0,Y=0)
\eg
What is new is the observation that
several generations of quarks may be classified by means
of reducible but indecomposable representations of 
$su(2\vert 1)$, viz.
\[
\rho^{(q)}\lplus\rho^{(q)}\quad\mbox{or}
\quad\rho^{(q)}\lplus\rho^{(q)}\lplus\rho^{(q)}\, ,
\]
where $\rho^{(q)}$ is given in eq. (21). 
These representations which offer a natural place for 
generation mixing,
when supplemented by a natural physical assumption,
lead to interesting relations between quark masses and 
CKM matrix
elements.

Likewise, for leptons and massive neutrinos one would 
use
\[
\left(\rho^{(\ell)}\lplus\rho_0^{(\ell)}\right)\lplus
\left(\rho^{(\ell)}\lplus\rho_0^{(\ell)}\right)\,\mbox{ 
or }\,
\]
\[
\left(\rho^{(\ell)}\lplus\rho_0^{(\ell)}\right)\lplus
\left(\rho^{(\ell)}\lplus\rho_0^{(\ell)}\right)\lplus
\left(\rho^{(\ell)}\lplus\rho_0^{(\ell)}\right)
\]
so as to obtain relations
between lepton masses and neutrino
mixing matrices \cite{CES,HS}. Without going into the 
details here
let me just state the additional,
physical assumption and quote a typical result.
The assumption is that in the absence of electroweak 
interactions
the masses in each charge sector are equal. In the case 
of {\em two\/}
generations this gives a parameter-free relation, to 
witness
\begin{itemize}
\item[(a)] For quarks,
\[
\cos\theta_C=\frac{\sqrt{m_um_d}+\sqrt{m_cm_s}}
                  {\sqrt{(m_u+m_c)(m_d+m_s)}}\, ;
\]
\item[(b)] For leptons,
\[
\cos\theta=\frac{\sqrt{m_1m_e}+\sqrt{m_2m_\mu}}
                  {\sqrt{(m_1+m_2)(m_e+m_\mu)}}\, ,
\]
($m_1$ and $m_2$ being the masses of the neutrinos).
\end{itemize}
In the case of {\em three\/} generations of quarks we 
find a pattern
of the mass matrices for {\em up-\/} and {\em 
down-\/}type quarks,
as well as for the CKM mixing matrix which is very 
similar
to purely phenomenological analyses of CKM mixing
(for references see \cite{HS}).
This is the more remarkable as the phenomenological 
approaches
start from a physical assumption which is just the 
opposite of ours:
they suppose that initially only one generation is 
massive and,
in fact, very heavy, the
other two are essentially massless, the masses of light 
quarks being
generated by mixing matrix elements. The analysis within 
the
Mainz-Marseille model not only reproduces the correct 
mixing pattern,
it also yields analytic expressions for the CKM matrix 
in terms of
quark masses and a few parameters which are useful for 
further
studies.

An analysis of leptonic mass matrices along the same 
lines is in
progress \cite{HPaS}. The results we have obtained so 
far look
promising and are compatible with some of the possible 
evidence
for nonzero neutrino masses and mixing.\\[4pt]

\noindent
{\bf 4. Possible parameter relations and some 
conclusions}\\

At the classical level, the Connes-Lott construction of 
the standard
model leads to relations between $m_H$, the mass of the 
Higgs particle,
and $m_W$, the mass of the $W$, in the form of rather 
tight
inequalities and via the fermionic mass matrix (3) 
which, of course,
is dominated by the top quark mass. These relations were 
studied in
detail and were especially advocated by one Marseille 
group \cite{Kas}.
In a recent paper Chamseddine and Connes \cite{ChC} 
these relations are
interpreted as being valid at a grand unification scale 
and are continued
by renormalization group equations to the low-energy 
regime. Wulkenhaar,
finally, who makes use of a Connes-Lott framework 
restricting the algebra
${\cal A}_M$ to be a Lie algebra, obtains somewhat 
different, but
comparable relations \cite{Wul}.

These relations obviously need some comment. A first 
criticism that is
often put forward concerns quantization. According to 
our present
understanding the construction yields a {\em 
classical\/} Lagrangian
which is the one of the standard model (or is very close 
to it) and which,
in a second step, has to be quantized following the 
usual procedures.
Then, as there is no new symmetry in the model, there 
does not seem
to be any obvious mechanism that would protect 
classical,
tree-level relations after quantization.

The second criticism involves a more technical point: 
Obviously,
in deriving the action a technically rather complicated 
scalar product
is involved. Let me illustrate this point by the example 
of the
Mainz-Marseille model where matters are more transparent 
in this respect. 
The calculation of the Lagrangian from the square of the 
curvature leads
to traces (over matrices and over Lie algebras) as well 
as to scalar
products of exterior forms of grade 0, 1, and 2. As is 
well known
the latter have no canonical normalization and, 
therefore, the terms
in the Lagrangian which stem from these scalar products 
will have
arbitrary relative strengths. Another way of saying this 
is that
there will be no canonical relation between, say, $m_H$ 
and $m_W$.
If, in turn, one decided to adopt the simplest and most 
natural
choice for these scalar products, one would fix that 
mass ratio,
obtaining e.g. $m_H=\sqrt{2}m_W$ \cite{CES}. In the 
framework of
Connes and Lott some authors claim having chosen the 
most general
scalar product compatible with the requirements imposed 
on the
theory. This is presently being investigated by Paschke 
et al.
\cite{Pas} who study more carefully these requirements 
and who
question whether the most general scalar product has 
really been
used. The finding is that with a more general product 
still
compatible with the postulates of
the theory, there is essentially no restriction
on $m_W$, while $m_H$ has an upper bound in terms of 
$m_t$, the
top mass (but no nontrivial lower
bound), which is numerically similar to the results of 
\cite{Kas}.
This would imply, in practice, that it would not be 
possible to
fix the Higgs mass in terms of either $m_t$ or $m_W$, at 
the
classical level. Clearly, this problem needs further 
investigation.

Let me summarize by the following comments and 
conclusions.

Connes' construction of noncommutative manifolds is 
mathematically profound
and physically more ambitious than other models such as 
Mainz-Marseille. 
By replacing the study of spaces by the study of 
algebras it enables one
to investigate more general, ''noncommutative'' 
Riemannian manifolds
\cite{ConA}. In particular, gravity is included in an 
interesting way.
Its application to the eminently successful standard 
model of strong,
electromagnetic and weak forces, on the other hand, 
still has some problems
to be solved some of which may be of technical nature, 
while others might
need more thinking: the
fixing of the $U(1)$ quantum numbers of quarks and 
leptons is not natural
and somewhat murky. Yet, this issue is important not 
only for obtaining
the correct couplings but also for the dicussion of 
chiral anomalies.
The theory is formulated in Euclidean space-time and 
there are some
problems in continuing it to Minkowski signature. There 
is a problem
with fermion doubling whereby unphysical degrees of 
freedom are introduced,
similarly to what happens in lattice gauge theories, 
\cite{LiM}. Finally,
there is the issue of possible parameter relations that 
we mentioned above.

The Mainz-Marseille construction is much simpler but 
also less ambitious.
It yields the correct Lagrangian of the standard model, 
without the
potential freedom of choices in ordinary Yang-Mills 
theories. 
It has no problems with the $U(1)$ quantum numbers 
because these are fixed 
correctly by the classification of quarks and leptons. 
As a consequence, 
anomalies are absent from the start \cite{FSA}. 
Unless additional ad hoc assumptions
are introduced the model does not predict classical 
parameter relations.
On the other hand it provides an attractive framework 
for fermionic state
mixing in terms of mass matrices, with remarkable 
cross-relationships to 
phenomenology. Finally, in this approach, the geometric 
nature of the
Higgs phenomenon is found to be compelling and is 
particularly transparent.

More generally speaking, algebraic Yang-Mills-Higgs 
theories, independently
of which avenue one follows, provide more constraints 
and allow for less
freedom in constructing gauge theories. The constraints 
could fail when 
compared to experiment and, therefore, render these 
theories vulnerable.
It so happens that the constraints and the additional, 
noncommutative
structure agree with our empirical information on the 
fundamental
interactions.

\end{document}